\documentclass{article}
\usepackage{waspaa23,amsmath,graphicx,url,times}
\usepackage{color}
\usepackage{booktabs, multirow}

\usepackage{ifthen}
\title{Analysis of XLS-R for Speech Quality Assessment}

\name{Bastiaan Tamm,$^{1,2}$\sthanks{This research was supported by KU Leuven Special Research Fund grant C24M/22/025 and the Flemish Government under Onderzoeksprogramma AI Vlaanderen.}
      Rik Vandenberghe,$^{1}$
      Hugo Van hamme,$^{2}$}
\address{$^1$ Laboratory for Cognitive Neurology (LCN), Department of Neurosciences, KU Leuven, Belgium\\
         $^2$ Processing Speech and Images (PSI), Department of Electrical Engineering, KU Leuven, Belgium
}

\begin{document}

\ninept
\maketitle

\begin{sloppy}

\begin{abstract}
  In online conferencing applications, estimating the perceived quality of an audio signal is crucial to ensure high quality of experience for the end user. The most reliable way to assess the quality of a speech signal is through human judgments in the form of the mean opinion score (MOS) metric. However, such an approach is labor intensive and not feasible for large-scale applications. The focus has therefore shifted towards automated speech quality assessment through end-to-end training of deep neural networks. Recently, it was shown that leveraging pre-trained wav2vec-based XLS-R embeddings leads to state-of-the-art performance for the task of speech quality prediction. In this paper, we perform an in-depth analysis of the pre-trained model. First, we analyze the performance of embeddings extracted from each layer of XLS-R and also for each size of the model (300M, 1B, 2B parameters). Surprisingly, we find two optimal regions for feature extraction: one in the lower-level features and one in the high-level features. Next, we investigate the reason for the two distinct optima. We hypothesize that the lower-level features capture characteristics of noise and room acoustics, whereas the high-level features focus on speech content and intelligibility. To investigate this, we analyze the sensitivity of the MOS predictions with respect to different levels of corruption in each category. Afterwards, we try fusing the two optimal feature depths to determine if they contain complementary information for MOS prediction. Finally, we compare the performance of the proposed models and assess the generalizability of the models on unseen datasets.
\end{abstract}


\section{Introduction}
\label{sec:intro}

Given the increased dependence on online conferencing applications in recent years, the demand for a reliable automated method to assess perceived speech quality has grown. Common factors that can degrade conversational quality include jitter, latency, echo, packet loss, and distortion~\cite{grancharov_speech_2008}. The ground truth for perceived speech quality is derived from human judgments, usually in the form of Absolute Category Ratings (ACR). These ratings are used to calculate the mean opinion score (MOS), which is used as the ground truth for the perceived speech quality of a given audio sample. However, the collection of human judgments is extremely time- and labor-intensive, making it impractical for large-scale evaluations of speech quality. Many objective metrics such as the speech-to-reverberation modulation energy ratio do not necessarily correlate with the perceived speech quality~\cite{avila_non-intrusive_2019}. Efforts have therefore been dedicated towards machine learning approaches for speech quality assessment~\cite{leng_mbnet_2021, tseng_utilizing_2021}, for example based on long short-term memory (LSTM) networks~\cite{cauchi_non-intrusive_2019}. In recent work on the ConferencingSpeech 2022 challenge~\cite{yi22b_interspeech}, it was shown that leveraging pre-trained wav2vec-based XLS-R~\cite{babu22_interspeech} embeddings leads to state-of-the-art performance for the MOS prediction task~\cite{tamm22_interspeech}. Most notably, the model performed exceptionally well on the unseen TUB dataset, outperforming the next-closest competitor by 27.4\% and the overall second-place model by 42.9\% for the RMSE metric~\cite{yi22b_interspeech, tamm22_interspeech}. Whether the embeddings generalize well to other unseen datasets has not yet been investigated.

This paper aims to perform an in-depth analysis of the wav2vec-based XLS-R model for the task of speech quality assessment. We first note that the original paper~\cite{tamm22_interspeech} simply used the final hidden layer of the pre-trained 300M parameter XLS-R to train a MOS-prediction model instead of determining a feature depth that is optimal for the downstream task. We will therefore analyze the performance of embeddings extracted from each layer of XLS-R and also for each size of the pre-trained model (300M, 1B, 2B parameters). Also, we aim to validate the performance of the model on other unseen datasets.

\begin{figure}[t]
  \centering
  \centerline{\includegraphics[width=\columnwidth]{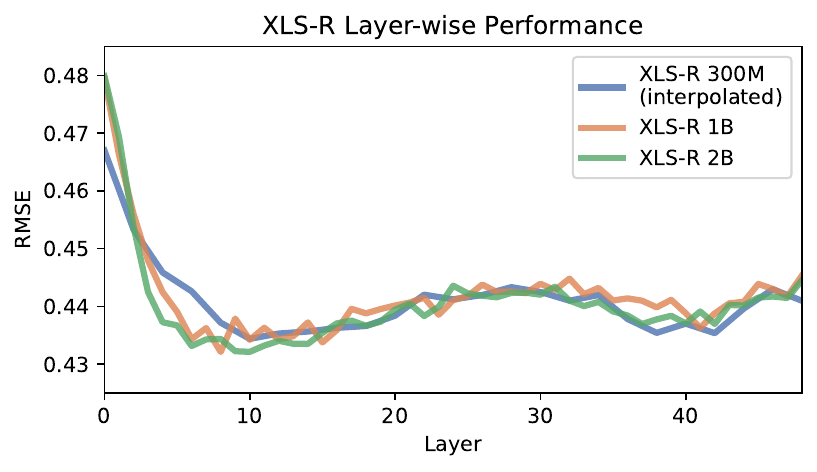}}
  \caption{Layer-wise performance of pre-trained XLS-R models on speech quality assessment task. The performance of each layer's activations is plotted for the three model sizes. This is measured using the best validation RMSE from all model configurations.
  This analysis was done on 35\% of the full dataset.
  }
  \label{fig:xlsr_layerwise_performance_fullds}
\end{figure}

\section{Experimental Setup}

\subsection{Datasets}
We use the same four corpora as the ConferencingSpeech 2022 challenge~\cite{yi22b_interspeech}, specifically
\begin{itemize}
    \item the \textbf{Tencent corpus}, consisting of around 14,000 Chinese speech clips with and without reverberation, ranging from 5 to 13.5 seconds long;
    \item the \textbf{Public switched telephone network (PSTN) corpus}, consisting of 80,000 English speech samples based on LibriVox with a length of 10 seconds each, containing both clean samples as well as samples with artificial background noise;
    \item the \textbf{Non-intrusive speech quality assessment (NISQA) corpus}~\cite{mittag_nisqa_2021}, a collection of more than 14,000 English and German speech clips with real as well as simulated noise;
    \item the \textbf{IU Bloomington (IUB)} corpus, comprising 36,000 English speech samples from the VOiCES~\cite{richey_voices_2018} and COSINE~\cite{stupakov_cosine_2009} datasets, ranging from three to six seconds long.
\end{itemize}
Each corpus is labeled with MOS values in a rating range of 1--5, which are derived from ACRs based on the International Telecommunication Union Telecommunication Standardization Sector (ITU-T) recommendation P.808~\cite{naderi_open_2020}. The only exception is the IU~Bloomington corpus, which follows ITU-R BS.1534~\cite{series2014method} and has a range of 0--100. These are converted to the range 1--5 for the experiments. Additionally, the count and standard deviation of the ACRs is provided for each audio sample in all corpora except for Tencent.

\subsection{Dataset Division} \label{se:dataset_division}

We use the same approach to dataset division as~\cite{tamm22_interspeech}. We define the challenge subset as the combination of the Tencent and PSTN corpora and the full dataset as the combination of all four corpora. The full training and validation sets are constructed by shuffling the samples in the full dataset and using 85\% for training and 15\% for validation. Finally, the training and validation subsets are constructed by keeping only the Tencent and PSTN samples from the original training and validation sets. We define an additional dataset, which will be referred to as the unseen dataset, which only contains the samples from NISQA and IUB. Thus, the models will be trained on the challenge subset and evaluated on the ``unseen" datasets.

\subsection{Model Architecture}

The model takes as input a sequence of 384 extracted features, which can be pre-trained XLS-R embeddings from a specific layer or MFCC features for comparison. The features are extracted as a preprocessing step and are not finetuned. The speech quality prediction model consists of three modules: a linear down-projection to the size of the hidden space, a bidirectional LSTM (Bi-LSTM) or transformer module to model temporal dependency, and an attention-based pooling module~\cite{mittag_nisqa_2021} to map to the output space. 

The number of Bi-LSTM or transformer layers as well as the hidden size are varied across models. The best transformer models use a hidden size of 32, 4 attention heads and 4 layers; the best Bi-LSTM models use a hidden size of 32 in each direction and 2 layers. We apply batch normalization at the input and after the Bi-LSTM/transformer. Subsequently, the outputs from the attention pooling module are mapped to the range (0,1) with a sigmoid function. These final outputs are referred to as normalized MOS values. This intermediate normalization step ensures that the output range of the attention pooling layer is unrestricted. During the evaluation of the model, the normalized MOS predictions are mapped to the original 1--5 MOS range.

\subsection{Training Details}

For MFCC calculation, we use the implementation by torchaudio~\cite{yang_torchaudio_2022} with the default parameters and a sample rate of 16~kHz. The XLS-R feature extraction uses the \textit{facebook/wav2vec2-xls-r-\{300m,1b,2b\}} models available on HuggingFace~\cite{wolf_transformers_2020}.

The models are implemented using the PyTorch (v.1.11.0)
and PyTorch Lightning (v.1.8.6)
libraries in Python 3.9. Training is performed using the PyTorch Lightning trainer. The network is trained using the ADAM optimizer~\cite{kingma2014adam}, a learning rate of $3\times10^{-3}$, batch size of 60, and  MSE loss. Each model is trained for a total of 30 epochs, and the model with the lowest validation loss is selected.

\section{XLS-R Layer-wise Performance}

First, we look at the performance of embeddings extracted from each layer of XLS-R and also for each size of the model (300M, 1B, 2B parameters). The results of the training are shown in Figure~\ref{fig:xlsr_layerwise_performance_fullds}. On the horizontal axis, the layer from which the activations are taken is displayed. This includes the output of the CNN (layer~0) and all transformer layers (layer~1--48). We interpolate the results of XLS-R 300M for this visualization since this model only has 24 transformer layers. On the vertical axis, we display the performance of the downstream speech quality prediction model on the validation split of the full dataset (RMSE metric, lower is better).

We hypothesized that the performance would rapidly improve over the first few layers and reach an optimum in the lower- or mid-level features, where room acoustics and noise characteristics are best modeled. Then, we believed that the performance would gradually degrade as the layer index further increased, as the highly contextualized speech representations would probably be less suitable to detect localized sources of speech degradation.

The first part of the hypothesis appears to be validated across the three model sizes, as the models reach an optimum around layer~10 (layer~5 for XLS-R 300M). Surprisingly, there seems to be a second local optimum around layer~41 (layer~21 for XLS\mbox{-}R 300M). It appears that there is a certain level of contextualization that is beneficial for speech quality assessment. In the following sections, we will investigate the properties of the low-level and high-level XLS-R embeddings.

\section{Corruption Sensitivity Analysis}

\begin{figure*}[t!]
  \centering
  \centerline{\includegraphics[width=\textwidth]{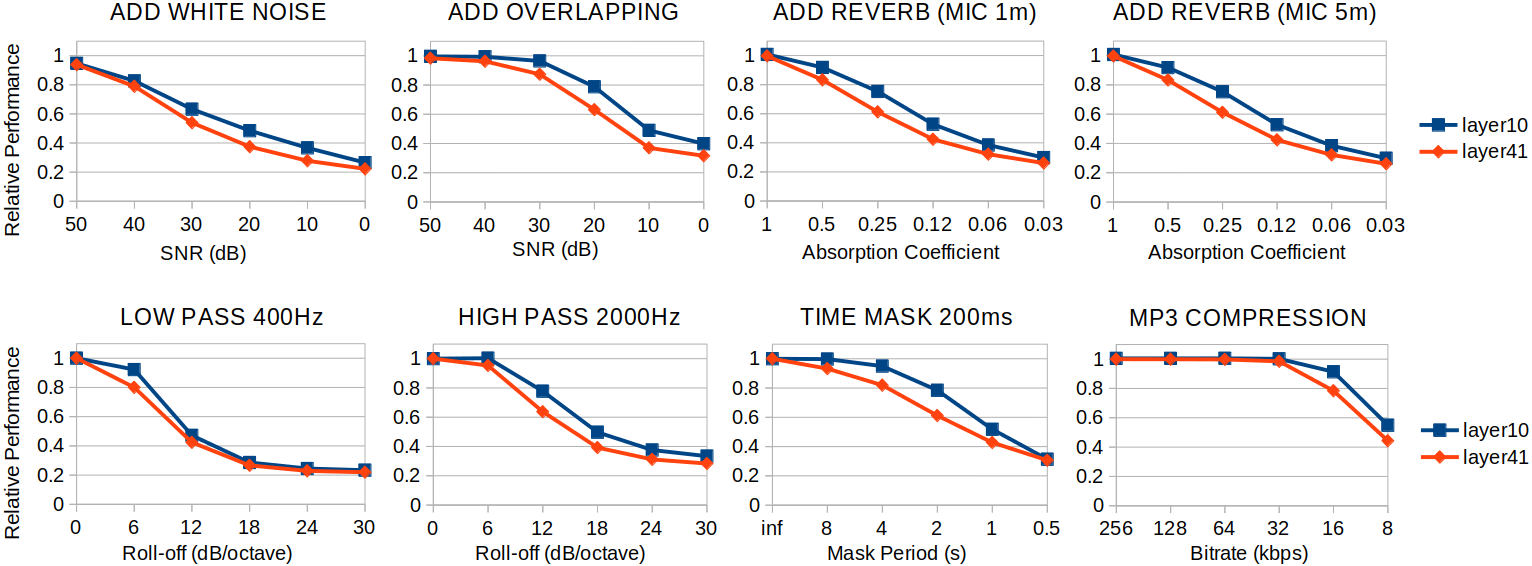}}
  \caption{Corruption sensitivity analysis of 2B-parameter model for full dataset.}
  \label{fig:results_sensitivity_2b_full}
\end{figure*}

Next, we investigate the reason for the two distinct optima in Figure~\ref{fig:xlsr_layerwise_performance_fullds}. We hypothesize that the lower-level features better model the more typical conditions that affect the quality of audio in online applications (e.g., room acoustics, echo, packet loss, distortion) and that the higher-level features capture some sort of speech content and intelligibility. Moreover, we expect the two levels of representation to have complementary information and that a fusion model will outperform each feature level individually. This section focuses on the first hypothesis, while Section~\ref{sec:model_comparison} addresses the latter.

To investigate what types of information are represented at each level, we artificially inject different types of noise and corruption and observe what effect this has on the predictions. We test a variety of corruption techniques: white Gaussian noise, overlapping speech, simulated reverb, low-/high-pass filter, time masking, and MP3 compression. The goal is to determine if a particular feature depth is more sensitive to certain types of degradation. The implementation is done using the audiomentations\footnote{https://github.com/iver56/audiomentations} Python package.

 
To estimate the sensitivity, we keep the ground-truth labels fixed and calculate the RMSE between the predictions of the corrupted audio and the ground-truth labels. Naturally, as the level of corruption increases, the predictions will be affected and the RMSE will increase. For visualization, we plot the relative performance of the corrupted predictions compared to the original predictions. A constant value of 1 indicates that the model is insensitive to the injected corruption, whereas a steep negative slope means the predictions are highly sensitive to the corruption.

The results are shown in Figure~\ref{fig:results_sensitivity_2b_full}. It can be seen that the model based on high-level XLS-R embeddings (orange line) is more sensitive to all types of corruption, a finding which does not directly support the hypothesis. It seems that perturbations in a given hidden layer of XLS-R propagate and may slightly magnify in later layers. This result can most likely be attributed to the fact that the wav2vec2 training procedure is not designed to generate embeddings that are insensitive to degraded audio. During pre-training, the model is trained to predict masked quantized representations derived from raw audio. The only robustness we would expect is to masking in the quantized representation space. Apparently, this does not translate to insensitivity to time masking, but this could also be due to the relatively long mask window.


This gives an idea of why the XLS-R embeddings are useful for speech quality assessment in the first place. We expect that a noise-robust version of wav2vec2, such as those proposed in~\cite{wang_wav2vec_switch,zhu_wav2vec2_noise_robust,wang_wav2vec2_noise_robust}, would not be useful for speech quality assessment. We have seen in our experiments, for example, that the version of XLS\mbox{-}R finetuned on multilingual speech translation\footnote{https://huggingface.co/facebook/wav2vec2-xls-r-2b-22-to-16} performs very poorly when embeddings are extracted from the final hidden layer, presumably because these embeddings focus on speech content and are more invariant to noise characteristics. 

\section{Model Comparison}
\label{sec:model_comparison}

\begin{figure}[t]
  \centering
  \centerline{\includegraphics[width=\linewidth]{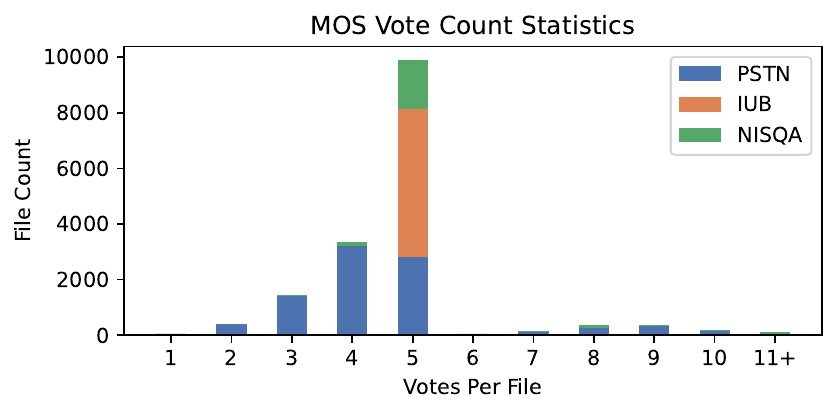}}
  \caption{MOS vote count statistics for PSTN, IUB and NISQA validation sets. No counts were available for the Tencent corpus.}
  \label{fig:mos_votes}
\end{figure}

\begin{table*}[t!]
\centering
\footnotesize
\begin{tabular}{ccc|cc|cc|ccc}
\toprule
\textbf{Group} & \textbf{Train Set} &
\textbf{XLS-R Layer}
& \textbf{Tencent} & \textbf{PSTN} & \textbf{NISQA} & \textbf{IUB} & \textbf{Subset} & \textbf{Unseen} & \textbf{Full} \\
\midrule
\textbf{Baseline \cite{tamm22_interspeech}} & subset & 24 & \textbf{0.3037} & \textbf{0.5022} & \textbf{0.5907} & \textbf{0.5067} & \textbf{0.4759} & \textbf{0.5323} & \begin{tabular}{c}{\tiny \textit{(0.5000)}}\end{tabular} \\
\textbf{DNSMOS \cite{reddy2021dnsmos}} & / & / &  \begin{tabular}{c}{\tiny \textit{(0.8338)}}\end{tabular} &  \begin{tabular}{c}{\tiny \textit{(0.7037)}}\end{tabular} & \textbf{0.8718} & \textbf{0.5452} &  \begin{tabular}{c}{\tiny \textit{(0.7262)}}\end{tabular} & \textbf{0.6565} & \begin{tabular}{c}{\tiny \textit{(0.6982)}}\end{tabular} \\
\midrule
\multirow{2}{*}{\textbf{\begin{tabular}[c]{@{}c@{}}MFCC\\ Transformer\end{tabular}}} & full & / & 0.5932 & 0.5924 & \begin{tabular}{c}{\tiny \textit{(0.6734)}}\end{tabular} & \begin{tabular}{c}{\tiny \textit{(0.3854)}}\end{tabular} & 0.5925 & \begin{tabular}{c}{\tiny \textit{(0.4865)}}\end{tabular} & \textbf{0.5511} \\
\textbf{} & subset & / & 0.5762 & 0.5992 & 0.8280 & 0.7775 & 0.5955 & 0.7924 & \begin{tabular}{c}{\tiny \textit{(0.6840)}}\end{tabular} \\
\midrule
\multirow{6}{*}{\textbf{\begin{tabular}[c]{@{}c@{}}XLS-R 300M\\ Transformer\end{tabular}}} & full & 5 & 0.3340 & 0.5002 & \begin{tabular}{c}{\tiny \textit{(0.4251)}}\end{tabular} & \begin{tabular}{c}{\tiny \textit{(0.3711)}}\end{tabular} & 0.4774 & \begin{tabular}{c}{\tiny \textit{(0.3875)}}\end{tabular} & 0.4423 \\
\textbf{} & full & 21 & 0.3119 & \underline{\textbf{0.4953}} & \begin{tabular}{c}{\tiny \textit{(0.4310)}}\end{tabular} & \begin{tabular}{c}{\tiny \textit{(0.3758)}}\end{tabular} & \underline{\textbf{0.4706}} & \begin{tabular}{c}{\tiny \textit{(0.3925)}}\end{tabular} & 0.4400 \\
\textbf{} & full & 5+21 & 0.3115 & 0.4976 & \begin{tabular}{c}{\tiny \textit{(0.4148)}}\end{tabular} & \begin{tabular}{c}{\tiny \textit{(0.3768)}}\end{tabular} & 0.4726 & \begin{tabular}{c}{\tiny \textit{(0.3882)}}\end{tabular} & \textbf{0.4396} \\
\textbf{} & subset & 5 & 0.3212 & 0.5036 & 0.6256 & 0.5049 & 0.4790 & 0.5425 & \begin{tabular}{c}{\tiny \textit{(0.5063)}}\end{tabular} \\
\textbf{} & subset & 21 & 0.3003 & 0.5068 & 0.5694 & 0.5025 & 0.4796 & 0.5227 & \begin{tabular}{c}{\tiny \textit{(0.4979)}}\end{tabular} \\
\textbf{} & subset & 5+21 & \underline{\textbf{0.2948}} & 0.5055 & \textbf{0.5683} & \textbf{0.4886} & 0.4779 & \textbf{0.5129} & \begin{tabular}{c}{\tiny \textit{(0.4927)}}\end{tabular} \\
\midrule
\multirow{6}{*}{\textbf{\begin{tabular}[c]{@{}c@{}}XLS-R 1B\\ Transformer\end{tabular}}} & full & 10 & 0.3127 & \textbf{0.4988} & \begin{tabular}{c}{\tiny \textit{(0.4285)}}\end{tabular} & \begin{tabular}{c}{\tiny \textit{(0.3676)}}\end{tabular} & \textbf{0.4738} & \begin{tabular}{c}{\tiny \textit{(0.3862)}}\end{tabular} & \textbf{0.4396} \\
\textbf{} & full & 41 & \textbf{0.3014} & 0.5007 & \begin{tabular}{c}{\tiny \textit{(0.4389)}}\end{tabular} & \begin{tabular}{c}{\tiny \textit{(0.3689)}}\end{tabular} & 0.4743 & \begin{tabular}{c}{\tiny \textit{(0.3904)}}\end{tabular} & 0.4415 \\
\textbf{} & full & 10+41 & 0.3188 & 0.5021 & \begin{tabular}{c}{\tiny \textit{(0.4658)}}\end{tabular} & \begin{tabular}{c}{\tiny \textit{(0.3983)}}\end{tabular} & 0.4774 & \begin{tabular}{c}{\tiny \textit{(0.4189)}}\end{tabular} & 0.4541 \\
\textbf{} & subset & 10 & 0.3198 & 0.5126 & \underline{\textbf{0.5456}} & 0.5815 & 0.4868 & 0.5713 & \begin{tabular}{c}{\tiny \textit{(0.5235)}}\end{tabular} \\
\textbf{} & subset & 41 & 0.3168 & 0.5118 & 0.5657 & \underline{\textbf{0.4656}} & 0.4858 & \underline{\textbf{0.4966}} & \begin{tabular}{c}{\tiny \textit{(0.4903)}}\end{tabular} \\
\textbf{} & subset & 10+41 & 0.3380 & 0.5050 & 0.5748 & 0.5288 & 0.4821 & 0.5425 & \begin{tabular}{c}{\tiny \textit{(0.5080)}}\end{tabular}  \\
\midrule
\multirow{6}{*}{\textbf{\begin{tabular}[c]{@{}c@{}}XLS-R 2B\\ Transformer\end{tabular}}} & full & 10 & 0.3520 & 0.5139 & \begin{tabular}{c}{\tiny \textit{(0.4717)}}\end{tabular} & \begin{tabular}{c}{\tiny \textit{(0.3739)}}\end{tabular} & 0.4915 & \begin{tabular}{c}{\tiny \textit{(0.4046)}}\end{tabular} & 0.4575 \\
\textbf{} & full & 41 & 0.3236 & \textbf{0.4992} & \begin{tabular}{c}{\tiny \textit{(0.4297)}}\end{tabular} & \begin{tabular}{c}{\tiny \textit{(0.3813)}}\end{tabular} & \textbf{0.4754} & \begin{tabular}{c}{\tiny \textit{(0.3959)}}\end{tabular} & \textbf{0.4442} \\
\textbf{} & full & 10+41 & 0.3111 & 0.5037 & \begin{tabular}{c}{\tiny \textit{(0.4217)}}\end{tabular} & \begin{tabular}{c}{\tiny \textit{(0.3987)}}\end{tabular} & 0.4780 & \begin{tabular}{c}{\tiny \textit{(0.4055)}}\end{tabular} & 0.4494 \\
\textbf{} & subset & 10 & 0.3034 & 0.5175 & 0.6277 & 0.4899 & 0.4894 & 0.5334 & \begin{tabular}{c}{\tiny \textit{(0.5081)}}\end{tabular} \\
\textbf{} & subset & 41 & \textbf{0.2977} & 0.5054 & \textbf{0.5724} & 0.4897 & 0.4781 & 0.5150 & \begin{tabular}{c}{\tiny \textit{(0.4937)}}\end{tabular} \\
\textbf{} & subset & 10+41 & 0.3069 & 0.5031 & 0.6036 & \textbf{0.4743} & 0.4770 & \textbf{0.5150} & \begin{tabular}{c}{\tiny \textit{(0.4931)}}\end{tabular} \\
\midrule
{\textbf{Human}} & / & / & / &  \begin{tabular}{c}{\tiny \textit{(0.7889)}}\end{tabular} & \textbf{0.6738} & \textbf{0.6573} & / & \textbf{0.6629} & / \\
{without quantization} & / & / & / &  \begin{tabular}{c}{\tiny \textit{(0.7342)}}\end{tabular} & 0.6088 & 0.6571 & / & / & / \\
\bottomrule
\end{tabular}
\caption{Model comparison for each corpus individually and the challenge subset (Tencent+PSTN), unseen (NISQA+IUB), and full datasets. The metric is RMSE on the respective validation set, lower is better. Values of interest are shown in bold. The overall best model per column is underlined. Some values are not relevant for the discussion but are provided for completeness: these are displayed in a smaller font between parentheses. For example, models trained on the full dataset have technically seen the so-called ``unseen" dataset. Also, comparing DNSMOS, which has not been trained on the challenge subset, to models where this is the case would not be a fair comparison. The final rows display the ``RMSE" of the human annotations and the estimated human RMSE without integer limitation (modeled as uniform quantization noise).}
\label{tab:results_comparison}
\end{table*}

\begin{figure*}[t!]
  \centering
  \centerline{\includegraphics[width=\textwidth]{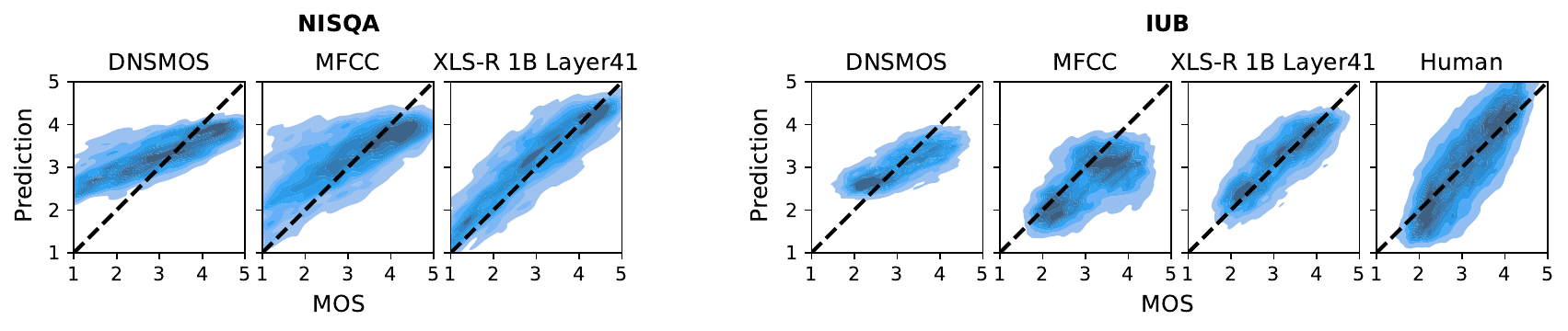}}
  \caption{Visualization of MOS predictions on unseen corpora. The human ACRs are also visualized for the IUB corpus.}
  \label{fig:prediction_visualization}
\end{figure*}

To assess if the two feature depths contain complementary information for speech quality assessment, we developed layer-fusion models and compared their performance to the single-layer models. We also included the performance of the baseline model from~\cite{tamm22_interspeech} and the popular DNSMOS~\cite{reddy2021dnsmos} model for comparison. As a final metric for comparison, we calculate the RMSE of human annotations with respect to the mean opinion score. This can be derived for all corpora except for Tencent since the count and standard deviation of the ACRs are provided per audio sample. 

\begin{equation}
  \label{eqn:rmse_human}
  RMSE_{\mathrm{human}}=\sqrt{\frac{\sum_{j} s_j^2 \cdot (N_j-1)}{\sum_{j} N_j} }
\end{equation}

The expression $s_j^2 \cdot (N_j-1)$ is equal to the sum of squared errors for a given audio sample $j$ with respect to the MOS ($s_j$ is the Bessel-corrected standard deviation). We sum this expression over all samples $j$ to obtain the global sum of squared errors and divide by the total number of votes to obtain the mean squared error. This is followed by a square root operation to obtain the desired metric. Contrary to the model predictions, humans are generally restricted to integer scores, so this must be considered as an extra source of variance.
Finally, a histogram of the number of votes per audio sample is shown in Figure~\ref{fig:mos_votes}.


The model comparison results are shown in Table~\ref{tab:results_comparison}. Values of interest are shown in bold, and the overall best model per column is underlined as well. We achieve slightly better performance than the baseline on the challenge subset (1.1\% better). More importantly, we validate the claim that the XLS-R-based model indeed performs exceptionally well on unseen data~\cite{tamm22_interspeech}. We show that the baseline model achieves an RMSE of 0.5323 on the validation set of the unseen NISQA+IUB corpora. This already outperforms DNSMOS (0.6565) and the RMSE of human annotations (0.6629). XLS\mbox{-}R 1B Layer41 performs even better with an RMSE of 0.4966 (6.7 / 24.4 / 25.1 \% better than baseline / DNSMOS / human respectively). Figure~\ref{fig:prediction_visualization} shows a visualization of the model predictions compared to DNSMOS and the MFCC model. Regarding layer fusion, we do not see a consistent improvement by applying early fusion to the two feature depths (weighted sum of inputs). The model attends to both inputs with weights -0.75/0.10 for layers 10/41 respectively.

\section{Conclusion}
In this paper, we have performed an analysis of the pre-trained XLS-R models for the task of speech quality assessment. We found that using specific layer activations results in improved performance compared to using the final hidden layer. Specifically, there are two local optima for feature depth selection around layers 10 and 41 (layers 5 and 21 for XLS-R 300M); however, the reason for the two distinct optima is still unclear. Finally, we showed that the proposed models\footnote{Models: {\scriptsize \url{https://github.com/lcn-kul/xls-r-analysis-sqa}}} substantially outperform DNSMOS and have lower variance than human annotators.

\bibliographystyle{IEEEtran}
\bibliography{refs23}

\begin{thebibliography}{10}
\providecommand{\url}[1]{#1}
\def\UrlFont{\rmfamily}
\providecommand{\newblock}{\relax}
\providecommand{\bibinfo}[2]{#2}
\providecommand\BIBentrySTDinterwordspacing{\spaceskip=0pt\relax}
\providecommand\BIBentryALTinterwordstretchfactor{4}
\providecommand\BIBentryALTinterwordspacing{\spaceskip=\fontdimen2\font plus
\BIBentryALTinterwordstretchfactor\fontdimen3\font minus
  \fontdimen4\font\relax}
\providecommand\BIBforeignlanguage[2]{{%
\expandafter\ifx\csname l@#1\endcsname\relax
\typeout{** WARNING: IEEEtran.bst: No hyphenation pattern has been}%
\typeout{** loaded for the language `#1'. Using the pattern for}%
\typeout{** the default language instead.}%
\else
\language=\csname l@#1\endcsname
\fi
#2}}

\bibitem{grancharov_speech_2008}
V.~Grancharov and W.~Kleijn, ``\BIBforeignlanguage{en}{Speech {Quality}
  {Assessment}},'' in \emph{\BIBforeignlanguage{en}{Springer {Handbook} of
  {Speech} {Processing}}}.\hskip 1em plus 0.5em minus 0.4em\relax Springer,
  2008.

\bibitem{avila_non-intrusive_2019}
A.~R. Avila, H.~Gamper, C.~Reddy, R.~Cutler, I.~Tashev, and J.~Gehrke,
  ``Non-intrusive {Speech} {Quality} {Assessment} {Using} {Neural}
  {Networks},'' in \emph{2019 {IEEE} {International} {Conference} on
  {Acoustics}, {Speech} and {Signal} {Processing} ({ICASSP})}.\hskip 1em plus
  0.5em minus 0.4em\relax IEEE, 2019, pp. 631--635.

\bibitem{leng_mbnet_2021}
Y.~Leng, X.~Tan, S.~Zhao, F.~Soong, X.-Y. Li, and T.~Qin, ``{MBNET}: {MOS}
  {Prediction} for {Synthesized} {Speech} with {Mean}-{Bias} {Network},'' in
  \emph{2021 {IEEE} {International} {Conference} on {Acoustics}, {Speech} and
  {Signal} {Processing} ({ICASSP})}.\hskip 1em plus 0.5em minus 0.4em\relax
  IEEE, 2021, pp. 391--395.

\bibitem{tseng_utilizing_2021}
W.-C. Tseng, C.-y. Huang, W.-T. Kao, Y.~Y. Lin, and H.-y. Lee, ``Utilizing
  {Self}-{Supervised} {Representations} for {MOS} {Prediction},'' in
  \emph{Interspeech}, 2021, pp. 2781--2785.

\bibitem{cauchi_non-intrusive_2019}
B.~Cauchi, K.~Siedenburg, J.~F. Santos, T.~H. Falk, S.~Doclo, and S.~Goetze,
  ``Non-{Intrusive} {Speech} {Quality} {Prediction} {Using} {Modulation}
  {Energies} and {LSTM}-{Network},'' \emph{IEEE/ACM Transactions on Audio,
  Speech, and Language Processing}, vol.~27, no.~7, pp. 1151--1163, 2019.

\bibitem{yi22b_interspeech}
G.~Yi, W.~Xiao, Y.~Xiao, B.~Naderi, S.~Möller, W.~Wardah, G.~Mittag,
  R.~Culter, Z.~Zhang, D.~S. Williamson, F.~Chen, F.~Yang, and S.~Shang,
  ``{ConferencingSpeech 2022 Challenge: Non-intrusive Objective Speech Quality
  Assessment (NISQA) Challenge for Online Conferencing Applications},'' in
  \emph{Proc. Interspeech 2022}, 2022, pp. 3308--3312.

\bibitem{babu22_interspeech}
A.~Babu, C.~Wang, A.~Tjandra, K.~Lakhotia, Q.~Xu, N.~Goyal, K.~Singh, P.~{von
  Platen}, Y.~Saraf, J.~Pino, A.~Baevski, A.~Conneau, and M.~Auli, ``{XLS-R:
  Self-supervised Cross-lingual Speech Representation Learning at Scale},'' in
  \emph{Proc. Interspeech 2022}, 2022, pp. 2278--2282.

\bibitem{tamm22_interspeech}
B.~Tamm, H.~Balabin, R.~Vandenberghe, and H.~{Van hamme}, ``{Pre-trained Speech
  Representations as Feature Extractors for Speech Quality Assessment in Online
  Conferencing Applications},'' in \emph{Proc. Interspeech 2022}, 2022, pp.
  4083--4087.

\bibitem{mittag_nisqa_2021}
G.~Mittag, B.~Naderi, A.~Chehadi, and S.~Möller, ``{NISQA}: {A} {Deep}
  {CNN}-{Self}-{Attention} {Model} for {Multidimensional} {Speech} {Quality}
  {Prediction} with {Crowdsourced} {Datasets},'' in \emph{Interspeech}, 2021,
  pp. 2127--2131.

\bibitem{richey_voices_2018}
C.~Richey, M.~A. Barrios, Z.~Armstrong, C.~Bartels, H.~Franco, M.~Graciarena,
  A.~Lawson, M.~K. Nandwana, A.~Stauffer, J.~van Hout, P.~Gamble, J.~Hetherly,
  C.~Stephenson, and K.~Ni, ``Voices {Obscured} in {Complex} {Environmental}
  {Settings} ({VOICES}) corpus,'' in \emph{Interspeech}, 2018, pp. 1566--1570.

\bibitem{stupakov_cosine_2009}
A.~Stupakov, E.~Hanusa, J.~Bilmes, and D.~Fox, ``{COSINE} - {A} corpus of
  multi-party {COnversational} {Speech} {In} {Noisy} {Environments},'' in
  \emph{2009 {IEEE} {International} {Conference} on {Acoustics}, {Speech} and
  {Signal} {Processing} ({ICASSP})}.\hskip 1em plus 0.5em minus 0.4em\relax
  IEEE, 2009, pp. 4153--4156.

\bibitem{naderi_open_2020}
B.~Naderi and R.~Cutler, ``An {Open} source {Implementation} of {ITU}-{T}
  {Recommendation} {P}.808 with {Validation},'' in \emph{Interspeech}, 2020,
  pp. 2862--2866.

\bibitem{series2014method}
B.~Series, ``Method for the subjective assessment of intermediate quality level
  of audio systems,'' \emph{International Telecommunication Union
  Radiocommunication Assembly}, 2014.

\bibitem{yang_torchaudio_2022}
Y.-Y. Yang, M.~Hira, Z.~Ni, A.~Chourdia, A.~Astafurov, C.~Chen, C.-F. Yeh,
  C.~Puhrsch, D.~Pollack, D.~Genzel, D.~Greenberg, E.~Z. Yang, J.~Lian,
  J.~Mahadeokar, J.~Hwang, J.~Chen, P.~Goldsborough, P.~Roy, S.~Narenthiran,
  S.~Watanabe, S.~Chintala, V.~Quenneville-Bélair, and Y.~Shi, ``{TorchAudio}:
  {Building} {Blocks} for {Audio} and {Speech} {Processing},'' \emph{arxiv
  preprint arXiv:2110.15018}, 2022.

\bibitem{wolf_transformers_2020}
T.~Wolf, L.~Debut, V.~Sanh, J.~Chaumond, C.~Delangue, A.~Moi, P.~Cistac,
  T.~Rault, R.~Louf, M.~Funtowicz, J.~Davison, S.~Shleifer, P.~von Platen,
  C.~Ma, Y.~Jernite, J.~Plu, C.~Xu, T.~L. Scao, S.~Gugger, M.~Drame, Q.~Lhoest,
  and A.~Rush, ``Transformers: {State}-of-the-{Art} {Natural} {Language}
  {Processing},'' in \emph{Proceedings of the 2020 {Conference} on {Empirical}
  {Methods} in {Natural} {Language} {Processing} (EMNLP): {System}
  {Demonstrations}}, 2020.

\bibitem{kingma2014adam}
D.~P. Kingma and J.~Ba, ``Adam: A method for stochastic optimization,''
  \emph{arXiv preprint arXiv:1412.6980}, 2014.

\bibitem{wang_wav2vec_switch}
Y.~Wang, J.~Li, H.~Wang, Y.~Qian, C.~Wang, and Y.~Wu, ``Wav2vec-switch:
  Contrastive learning from original-noisy speech pairs for robust speech
  recognition,'' in \emph{ICASSP 2022 - 2022 IEEE International Conference on
  Acoustics, Speech and Signal Processing (ICASSP)}, 2022, pp. 7097--7101.

\bibitem{zhu_wav2vec2_noise_robust}
Q.-S. Zhu, J.~Zhang, Z.-Q. Zhang, M.-H. Wu, X.~Fang, and L.-R. Dai, ``A
  noise-robust self-supervised pre-training model based speech representation
  learning for automatic speech recognition,'' in \emph{ICASSP 2022 - 2022 IEEE
  International Conference on Acoustics, Speech and Signal Processing
  (ICASSP)}, 2022, pp. 3174--3178.

\bibitem{wang_wav2vec2_noise_robust}
H.~Wang, Y.~Qian, X.~Wang, Y.~Wang, C.~Wang, S.~Liu, T.~Yoshioka, J.~Li, and
  D.~Wang, ``Improving noise robustness of contrastive speech representation
  learning with speech reconstruction,'' in \emph{ICASSP 2022 - 2022 IEEE
  International Conference on Acoustics, Speech and Signal Processing
  (ICASSP)}, 2022, pp. 6062--6066.

\bibitem{reddy2021dnsmos}
C.~K. Reddy, V.~Gopal, and R.~Cutler, ``{DNSMOS}: A non-intrusive perceptual
  objective speech quality metric to evaluate noise suppressors,'' in
  \emph{ICASSP 2021 IEEE International Conference on Acoustics, Speech and
  Signal Processing (ICASSP)}.\hskip 1em plus 0.5em minus 0.4em\relax IEEE,
  2021, pp. 6493--6497.

\end{thebibliography}
%
%
%
%
%
%
%
%
%

\end{sloppy}
\end{document}